# Fast Thermal Relaxation in Cavity-Coupled Graphene Bolometers with a Johnson Noise Read-Out

Dmitri K. Efetov[1], Ren-Jye Shiue[2], Yuanda Gao[3], Brian Skinner[4], Evan D. Walsh[2], Hyeongrak Choi[2], Jiabao Zheng[2], Cheng Tan[3], Gabriele Grosso[2], Cheng Peng[2], James Hone[3], Kin Chung Fong[5] and Dirk Englund[2]

1. ICFO - Institut de Ciencies Fotoniques, The Barcelona Institute of Science and Technology, 08860 Castelldefels, Barcelona, Spain
2. Department of Electrical Engineering and Computer Science, Massachusetts Institute of Technology, Cambridge MA 02139
3. Department of Mechanical Engineering, Columbia University, New York NY 10027
4. Department of Physics, Massachusetts Institute of Technology, Cambridge MA 02139
5. Raytheon BBN Technologies, Quantum Information Processing Group, Cambridge, Massachusetts 02138, USA

**High sensitivity, fast response time and strong light absorption are the most important metrics for infrared sensing and imaging. The trade-off between these characteristics remains the primary challenge in bolometry. Graphene with its unique combination of a record small electronic heat capacity and a weak electron-phonon coupling has emerged as a sensitive bolometric medium that allows for high intrinsic bandwidths. Moreover, the materials light absorption can be enhanced to near unity by integration into photonic structures. Here, we introduce an integrated hot-electron bolometer based on Johnson noise readout of electrons in ultra-clean hexagonal boron nitride encapsulated graphene, which is critically coupled to incident radiation through a photonic nanocavity with Q=900. The device operates at telecom wavelengths and shows an enhanced bolometric response at charge neutrality. We report a noise equivalent power < 10 pW/Hz$^{1/2}$ , a record fast thermal relaxation time < 35ps and an improved light absorption at 5K. However the device can operate even above 300K with reduced sensitivity. We work out the performance mechanisms and limits of the graphene bolometer and give important insight towards the potential development of practical applications.**

Since the invention of the bolometer, its main design principles relied on efficient light absorption into a low-heat-capacity material and its exceptional thermal isolation from the environment. While reduced thermal coupling to its surroundings allows for an enhanced thermal response, it in turn strongly increases the thermal time constant and dramatically lowers the detector's bandwidth. Due to its record-low electronic heat capacity $C_e$ [1–3] and weak heat dissipation [3,4], absorbed radiation can give rise to an electron temperature $T_e$ that can be dramatically higher than the lattice temperature $T_s$, while maintaining a fast thermal relaxation time (Fig. 1(a)). These unique thermal properties and its broadband photon absorption from the uV to GHz frequencies [5] make graphene a promising platform for ultrasensitive and ultra-fast hot electron bolometers, calorimeters and single photon detectors for low energy photons [6–8].

Previous studies have demonstrated graphene's operation as a sensitive bolometer [9-11], however many challenges remain towards real applications. Here most experiments utilized graphene on a SiO$_2$ substrate with a slow DC transport readout. In order to achieve a high response, graphenes' weak temperature dependent resistance $\Delta R(T_e)/\Delta T_e$ had to be artificially increased by introducing disorder[12], patterning nanostructures[13], or opening a band gap in bilayer graphene[9]. As disorder introduces charge fluctuations at the CNP, it limits the detectors' response, as the lowest $C_e$ values in close vicinity to the CNP cannot be reached. Low light absorption of only ~2.3% (below THz wavelengths) [5,14,15] makes these device further impractical.

It was recently shown that it is possible to use a universal $T_e$ readout utilizing microwave frequency Johnson noise (JN) which is emitted by thermally agitated electrons in graphene [2,3,7,8,16]. This direct $T_e$ readout scheme does not require a large $\Delta R(T_e)/\Delta T_e$ and allows to use ultra-clean hexagonal boron nitride (hBN) encapsulated graphene devices without sacrificing the detectors sensitivity. In state-of-the-art devices the carrier density can be tuned to ultra-small values of n < 10$^{10}$ e/cm$^2$ [17] where $C_e$ can reach record low values of just one $k_b$. Moreover, recent progress in the critical coupling of graphene to resonant photonic cavities [18,19] and to antennas [8,13] showed that one can achieve close to unity light absorption into graphene for a broad range of frequencies.

In this manuscript we combine all of the above techniques in a proof-of-concept device. Fig. 1 (b) shows the schematics and the operation principle of the bolometer. A high quality hBN/G/hBN graphene heterostructure is integrated onto a suspended silicon photonic crystal cavity (PCC), with a linear-three-hole (L3) defect, which serves to form confined optical cavity modes. The resonant cavity mode of the PCC (Fig. 1 (c)) is critically coupled to the telecom spectrum and evanescently overlaps with the graphene sheet [18]. To maximize heat generation and to minimize heat dissipation, the graphene stack is etched into a circle with 5 $\mu$m diameter, which corresponds to the expected length scale over which the electron temperature relaxes back to thermal equilibrium [20]. It is further connected to gold leads with narrow, resistive graphene channels. In order to effectively transmit the microwave Johnson noise signal through the circuit, the high resistance of the graphene (R~20kΩ) is impedance matched to the microwave transmisison line (50Ω) by the addition of a carefully chosen inductor and capacitor. The so formed resonant RLC circuit has a resonance frequency of ~ 70MHz with a bandwidth of ~ 6MHz, and allows to transmit most of the JN power (JNP) (see SI) to a low noise amplifier (LNA) and a heterodyne circuit, which are situated at room temperature. The RF circuit forms the JN readout scheme, and its voltage output is directly proportional to the emitted JNP. By applying a gate voltage $\Delta V_{bg}$ (referenced to the CNP) to the PCC one can control the carrier density of the graphene sheet.

The device is probed at $T_s$=5K and $\Delta V_{bg}$=0V with linearly polarized excitation pulses from a tunable near-IR laser, which are normally incident onto the device through a cross-polarized confocal scanning setup [21] (SI). Simultaneously to spatial scanning of the laser, the calibrated JNP is read out [2]. As the JNP is directly proportional to $T_e$ (SI), we are able to obtain the electronic temperature increase due to laser excitation $\Delta T_e$ for each laser spot position. Fig. 1 (d) shows the optical image of the device and Fig. 1 (e) $\Delta T_e$ spatial map for 0$^0$ and 90$^0$ polarized light at the PCCs resonant frequency of 1532nm and an incident power of

$P_{in}$=16.5μW. Here, a sizeable $\Delta T_e$ increase is generated only when the laser is incident directly onto the graphene, confirming that the JN readout is only sensitive to thermal noise inside the graphene channel. As the PCCs resonant mode only couples to light with polarization along the x-axis, one can compare the response off (90⁰ polarization) and on (0⁰ polarization) resonance (Fig. 1 (e) left and middle). When off resonance, $\Delta T_e$ is roughly constant over the entire graphene region, assuming that 2.3% of the incident light is homogeneously absorbed in the whole graphene region. While when on resonance, in addition to direct light absorption, light is also resonantly coupled into the PCC. The resonant mode interacts with graphene and gives a strong rise to an enhancement of $\Delta T_e$ in the cavity region. With the thermal model presented in Fig. 1 (a) (SI), the theoretically expected $\Delta T_e$ map is calculated (Fig. 1 (e) right), and shows very good agreement with experimental findings.

To characterize the enhancement of light absorption by the PCC we perform measurements of the reflection spectrum of the irradiated cavity before the deposition of the hBN/G/hBN stack and after. Fig. 2 (a) shows the fundamental mode of the pure cavity with a resonance mode at 1512.8nm and a quality factor of $Q \sim 7000$. After the hBN/G/hBN deposition, the cavity resonance is red-shifted to 1531.7 nm and its $Q$ drops to $\sim 900$. Here the shift of the mode is mainly due to hBN's high dielectric constant of ~3.9, while its' broadening is caused by the light absorption in graphene. From coupled mode theory [18], we estimate that the total absorption of the cavity resonant field into graphene is ~ 45%.

Overall the enhanced light absorption strongly amplifies the photoresponse. Fig. 2 (b) shows $\Delta T_e$ as a function of incident laser wavelength for 0⁰ (on) and 90⁰ (off) polarized light with respect to the cavity x-axis (Fig. 1(c)). In the on state at a resonance wavelength of 1531.7nm, $\Delta T_e$ is enhanced by a factor of ~3 and follows a sinusoidal dependence as a function of the polarization angle (Fig. 2(b) inset). From the difference of the $\Delta T_e$ increase on and off resonance, we can deduce that the total absorption in graphene is enhanced from 2.3% to more than 7%. Since graphene absorbs ~ 45% of the cavity field, we extract a coupling efficiency of the cavity field to normal incidence laser light to be ~ 10%, consistent with simulation results. With improved fabrication however, one can expect near-unity absorption through critical coupling to a waveguide [22,23] or free-space [24,25], and experimentally >90% absorption from a photonic nanocavity into graphene has already been demonstrated [18].

We now probe $\Delta T_e$ as a function of absorbed laser power P (assuming 7% incident light absorption) at $T_s$=5K and a measurement time of $t_{in}$=1s (Fig. 3 (a)). Here the $\Delta T_e$(P) dependence transitions from a linear to a sublinear dependence with increased P, and is overall strongly reduced at elevated temperatures (inset Fig. 3(a)). We can model this behaviour with the 2D heat transfer equation below (Eq. 1), which takes into account the two main mechanisms : electronic dissipation through the WF law (first term) and through e-ph interactions (second term) :

$$\dot{q} = -\vec{\nabla}\cdot[\kappa\vec{\nabla}T_e] + \Sigma_{ep}[T_e^3 - T_s^3], (Eq. 1)$$

where $\Sigma_{ep}$ is the e-ph coupling strength, $\kappa=(\pi^2 k_b T_e)/3e\rho$ the thermal conductivity, and $\rho$ the device resistivity which is extracted from transport measurements (SI). Numerical modeling with only one fitting parameter $\Sigma_{ep}$=0.038 W/m$^2$K$^3$ produces a good match with the experiment and agrees well with previously reported values [3]. We can qualitatively understand the sub-linear behaviour by the overall enhancement of the electronic thermal conductivity $G_{th}$ at higher $T_s$, where both, WF and e-ph dissipation channels are strongly enhanced. The bolometric response due to laser heating is furthermore well consistent with the response through Joule heating of the sample with a large current [2] (Fig. 3(a)). Here minor deviations can be attributed to saturable light absorption in graphene at higher laser powers [26].

Ultimately, the bolometric response is limited by the electronic thermal conductivity $G_{th}$ through the simple relation $\Delta T_e$=P/$G_{th}$, which is a direct consequence of Eq. 1. From measurements at various $T_s$ with a fixed P=100nW, we obtain the $G_{th}(T_s)$ dependence (Fig. 3 (b)). It is in good agreement with the heat dissipation model, and follows a linear WF dependence at low $T_s$ and a power law e-ph dependence at higher $T_s$. Since both $C_e$ and $G_{th}$ are expected to be strongly reduced at the CNP [3], we test $\Delta T_e$ as a function of gate voltage $\Delta V_{bg}$. Fig. 3 (c) plots $\Delta T_e(\Delta V_{bg})$ for P=10nW and compares it to $R(\Delta V_{bg})$ (both taken at $T_s$=5K). The response sharply increases by a factor ~3 at the CNP and qualitatively follows the same dependence as $R(\Delta V_{bg})$. However the enhancement of the response at the CNP is predicted to be several orders of magnitude higher at mK temperatures, as thermal broadening of the Fermi energy is reduced and $C_e$ is dramatically lowered [2,8].

In analogy to previous work on graphene pn-junctions [27–29], we can now test the thermal relaxation time of the hot electrons $\tau$, which defines the intrinsic bandwidth limit of the detector. $\tau$ can be probed by the excitation of the bolometer with two $t_d$ time delayed pulses (SI), where the first pulse heats the electrons to an elevated $T_e$, and the second pulse heats the electron gas further. Because of the non-linear $\Delta T_e(P)$ dependence in Fig. 3 (a), the $T_e$ rise due to the second pulse is smaller than for the first when $t_d$ ~ 0ps and $T_e$ has not yet cooled down. This results in a greatly diminished integrated JNP signal. Fig. 4 (a) shows the normalized JNP as a function of $t_d$, JNP($t_d$)/JNP(120ps). The curves follow an inverse exponential form where the half-width-half-maximum (HWHM) of the central dip defines $\tau$, which is in very good agreement with previous studies [9,27–31]. As $G_{th}$ at low $T_s$ is dominated by WF, $\tau$ is in the order of only ~30ps, which is much faster than one would expect in e-ph limited device architectures [7].

From the above device characterization, we can define the bolometers performance characteristics, the noise equivalent power NEP and the intrinsic reset time, which is defined by $\tau$ (Fig. 4 (b)). Here the calibration of the bolometric readout scheme gives an overall sensitivity of $\delta T_e$~10mK for a measurement time of $t_{in}$=1s (SI), which together with the $G_{th}$ values from Fig. 3 (b) allows to extract the NEP values of the bolometer. The bolometer has no hard limit on the operation temperature and can be operated above liquid nitrogen and even room temperatures, with typical values reaching a NEP ~ 10 pW/Hz$^{1/2}$ and a $\tau$ ~ 30ps at $T_s$=5K and $\Delta V_{bg}$=0V. At the CNP, the NEP is strongly reduced, while $\tau$ only weakly varies with $\Delta V_{bg}$ [28,32]. Generally, while the NEP is dramatically reduced at low $T_s$, $\tau$ is increased.

As is given by the Dicke radiometer relation [33], the temperature sensitivity of the system is given by $\delta T_e = (T_e + T_{sys})/(Bt_{in})^{1/2}$, where B is the measurement bandwidth, $t_{in}$ the measurement time and $T_{sys}$ the system noise, which is dominated by the LNA (SI). From this relation it is clear that the sensitivity can be dramatically reduced by cooling the entire measurement circuit, the use of quantum limited amplifiers and by increasing B, and it was shown theoretically that the presented device concept could reach a NEP ~ $10^{-20}$ W/Hz$^{1/2}$ at $T_s$=10mK [8]. Given the ultra-fast intrinsic bandwidth of the bolometer, its ultimate detection speed is mainly limited by the readout electronics. While state-of-the-art LNAs can operate at frequencies above 10GHz, such a fast readout would dramatically lower $\delta T_e$ (as follows from the Dicke relation (SI)). In this regard, a practical graphene bolometer design, would have to operate at a carefully chosen trade-off space between the desired measurement sensitivity and a fast readout, which is captured by the value of the NEP. In addition, while the integration of the bolometer into resonant light structures will enhance light absorption, it will sacrifice graphenes' broadband absorption properties. It is however feasible that one could develop small bolometer arrays, similar to the recently developed microwave single photon detector arrays, where each bolometer is individually coupled to a PCC with a different resonant frequency, making the array sensitive to a broad range of frequencies.

In conclusion, the presented study shows the promises and challenges towards practical applications of graphene based bolometers. With its improved light absorption, high sensitivity, ultra-fast thermal relaxation time, no limitations on its operation temperature and the potential for mid-IR and THz applications, this proof-of-concept device provides unique characteristics for the development of advanced bolometers with new functionalities.

Acknowledgements:

We thank Leonid Levitov, Dan Prober and Frank Koppens for fruitful discussions. D.K.E. acknowledges support from the Ministry of Economy and Competitiveness of Spain through the "Severo Ochoa" program for Centres of Excellence in R&D (SEV-2015-0522), Fundació Privada Cellex, Fundació Privada Mir-Puig, the Generalitat de Catalunya through the CERCA program and the La Caixa Foundation. D.E. acknowledge support from the Office of Naval Research under grant number N00014-14-1-0349. Y.G., C.T. and J.H. acknowledge support from the US Office of Naval Research N00014-13-1-0662. K.C.F. acknowledges support from Raytheon BBN Technologies. B. S. was supported as part of the MIT Center for Excitonics, an Energy Frontier Research Center funded by the U.S. Department of Energy, Office of Science, Basic Energy Sciences under Award no. DE-SC0001088. J.Z. carried out research in part at the Center for Functional Nanomaterials, Brookhaven National Laboratory, which is supported by the U.S. Department of Energy, Office of Basic Energy Sciences, under Contract No. DE-SC0012704. This work is supported in part by the Semiconductor Research Corporation's NRI Center for Institute for Nanoelectronics Discovery and Exploration (INDEX).

Author contributions:

D.K.E., K.C.F. and D.E. conceived and designed the experiments: D.K.E. and R.J.S. performed the experiments: D.K.E. analyzed the data: B.S. performed the theoretical modeling of the data: Y.G., C. T., C.P., H.C., E.D.W., J.Z. and G.G. contributed materials/analysis tools: J.H., K.C.F., D.E. supported the experiments: D.K.E. wrote the paper.

Competing financial interests:

The authors declare no competing financial interests.

Competing financial and non-Financial interests:

The authors declare no competing financial and non-financial interests.

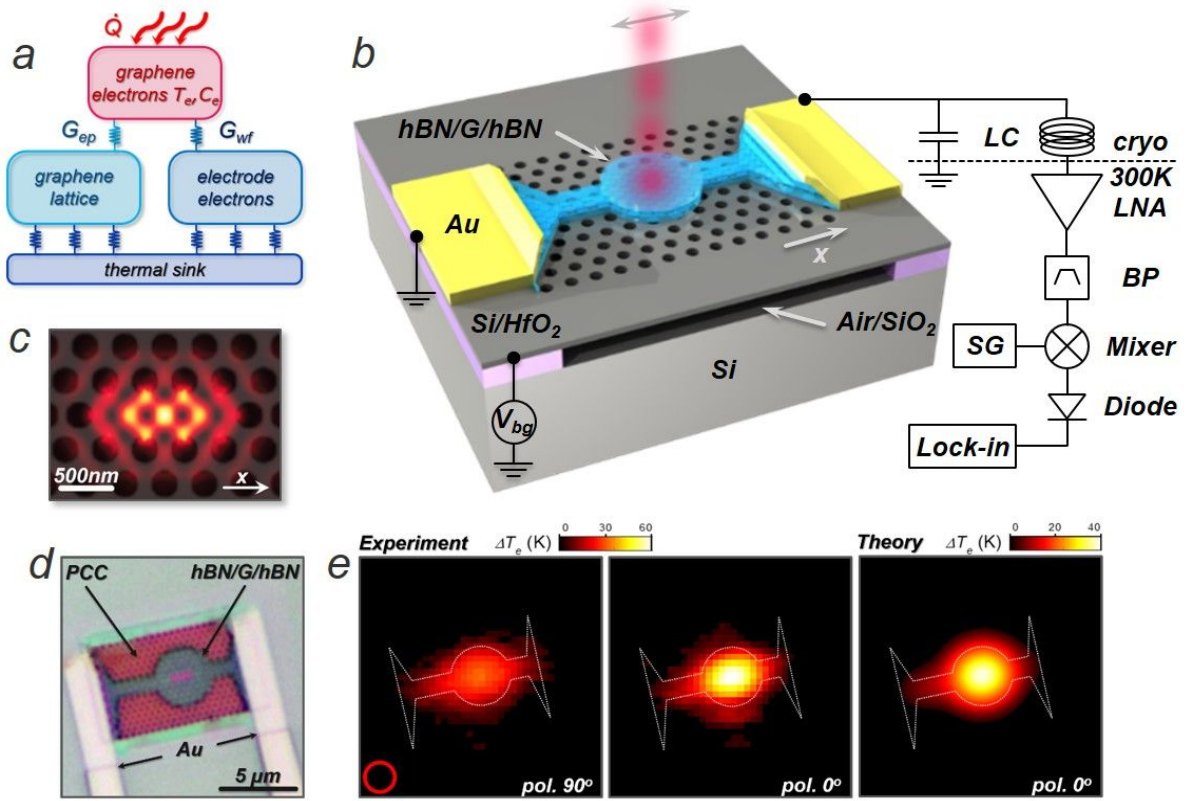

**Figure 1: Device schematics and operation principle.** (a) Schematics of heat dissipation channels of irradiated hot electrons in graphene. Primary mechanism includes the WF law ($G_{WF}$) and e-ph scattering ($G_{ep}$). (b) Schematic of the device, which consists of a side contacted hBN/G/hBN heterostructure on top of a suspended silicon L3 PCC. The device is impedance matched to a LC network at cryogenic temperatures and is read out by a heterodyne JN thermometry scheme at room temperature, which allows for an accurate reading of $T_e$. (c) Resonant modes form in the L3 PCC as is seen in the calculated spatial electric field intensity $|E|^2$ profile. (d) Optical microscope image of the device. (e) Map of the bolometric response of the device as a function of laser spot position (red circle – laser spot size) measured at $T_s$=5K and $\Delta V_{bg}$=0V. Left: $\Delta T_e$ for $90^0$ polarized light (PCC off resonance), middle: for $0^0$ polarized light (PCC on resonance) and right: theoretical model for $0^0$ polarized light. Overall the bolometric response occurs only when the laser beam is injected on the graphene covered area, and is strongly enhanced when the PCC mode is on resonance.

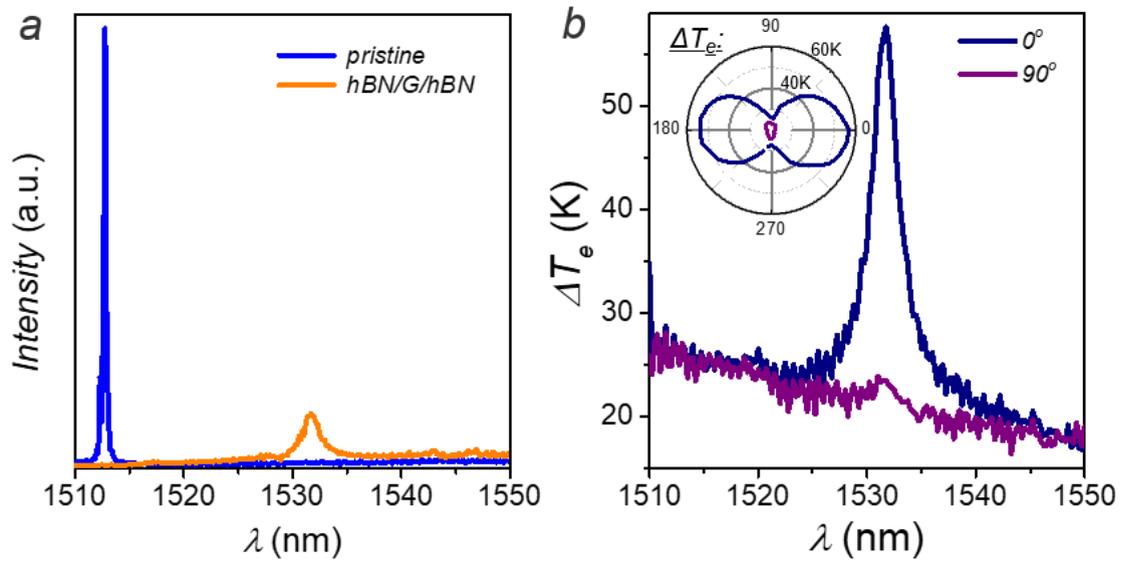

**Figure 2: Role of the PCC – enhancement of the bolometric response.** (a) Reflection spectra of the normally excited PCC cavity mode before and after the deposition of the hBN/G/hBN stack. Due to light absorption in the graphene and hBNs high dielectric constant the resonance is broadened and red-shifted. (b) Bolometric response as a function of the incident wavelength for $0^0$ (on) and $90^0$ (off) linearly polarized light, at $T_s$=5K and $\Delta V_{bg}$=0V. Inset shows sinusoidal polarization angle dependence of $\Delta T_e$ for different wavelengths on (blue - 1532.5nm) and off (purple - 1520nm) resonance.

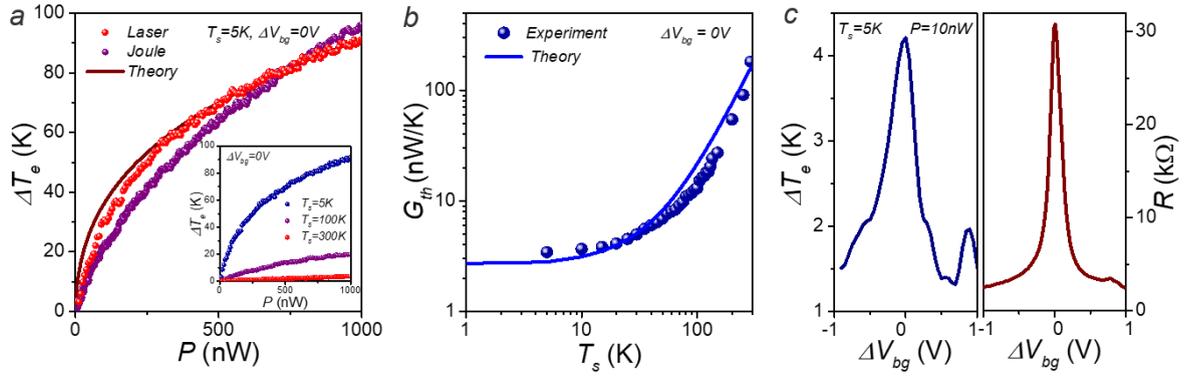

**Figure 3: Bolometric response.** (a) Bolometric response $\Delta T_e$ as a function of absorbed heating power P, for laser and Joule heating at $\Delta V_{bg}$=0V and $T_s$=5K. Inset shows $\Delta T_e$ for laser heating at various device temperatures $T_s$. (b) The extracted thermal conductance $G_{th}=P/\Delta T_e$ as a function of $T_s$. Overall the experimental results in (a) and (b) are consistent with a heat dissipation model based on the WF law and e-ph coupling (theory fits using Eq. 1). As the bolometric response is mainly limited by thermally activated dissipation mechanisms, it is strongly reduced at elevated $T_s$. (c) Gate dependence of $\Delta T_e(V_{bg})$ for P=10nW (left) and of $R(V_{bg})$ (right), both measured at $T_s$=5K. The curves show qualitatively similar dependence with a strong enhancement of $\Delta T_e$ at the CNP, due to a reduced $C_e$.

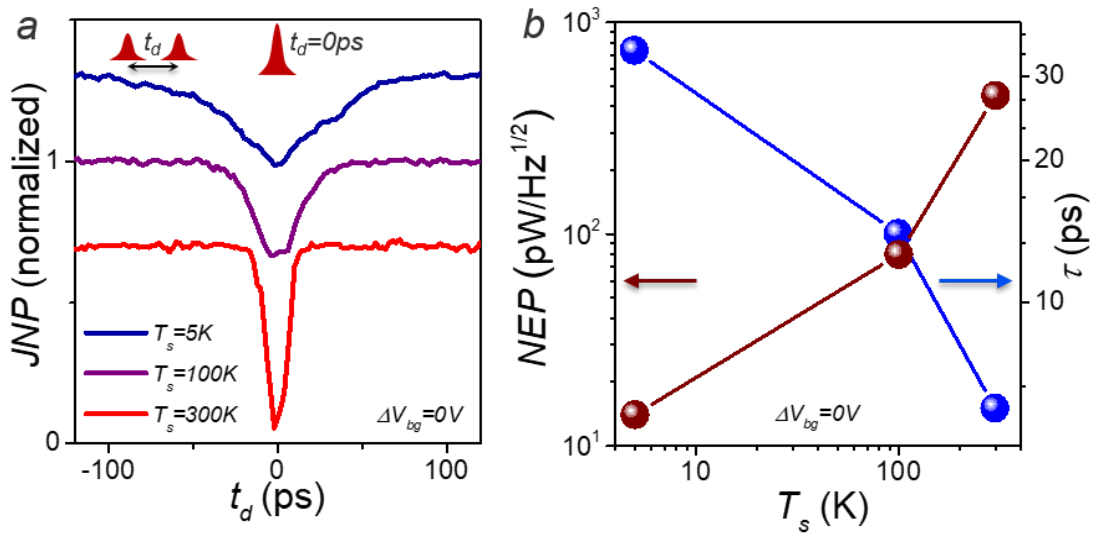

**Figure 4: Detector performance.** (a) Normalized Johnson noise power JNP($t_d$)/JNP(120ps) probed with two $t_d$ time delayed laser pulses at various device temperatures $T_s$ and at $V_{bg}$=0V (graphs are offset for clarity). From the HWHM of the dip around at $t_d$=0ps we extract the intrinsic thermal relaxation time of the bolometer $\tau$. As $G_{th}$ becomes reduced at low $T_s$, $\tau$ in turn increases. (b) NEP and $\tau$ as a function of $T_s$ at $\Delta V_{bg}$=0V. We extract a NEP ~ 10 pW/Hz$^{1/2}$ and an intrinsic reset time of $\tau$ ~ 30ps at $T_s$=5K, the bolometer has an unique combination of high read-out sensitivity and high intrinsic bandwidth. It has no hard limit on the operation temperature.

Data availability:

The data that support the findings of this study are available from the corresponding author upon reasonable and well-motivated request.

Additional information:

Supplementary information is available for this paper at …

Reprints and permissions information is available at www.nature.com/reprints. Correspondence and requests for materials should be addressed to D.K.E.

Publisher's note: Springer Nature remains neutral with regard to jurisdictional claims in published maps and institutional affiliations.

Methods:

Device fabrication: The hBN/graphene/hBN stacks were exfoliated and assembled using a van der Waals assembly technique and then transferred onto the top of the PCC cavity. The total thickness of the two BN layers is around 30 nm. We patterned the BN/Graphene/BN stack with hydrogen-silsesquioxane (HSQ) resist and CHF3+O2 plasma exposes the edges of graphene, which was subsequently contacted by Cr/Pd/Au (1/20/50 nm) metal leads using electron-beam evaporation. The L3 PCC cavities were fabricated on a silicon-on-insulator wafer using a series of electron-beam lithography, reactive ion etching, and a wet-etch undercut of the insulator to produce free-standing membranes. The top silicon membrane has a thickness of ~220 nm with a PCC lattice period of $a=440$ nm and an air hole radius $r=0.29a$. A linear-three-hole (L3) defect in the middle of the PCC lattice serves to form confined optical cavity modes.

Optical measurements: The devices were illuminated with a scanning cross-polarized confocal microscope setup. The reflection from the cavity was collect and coupled to a spectrometer to analyze its spectrum. For time-resolved measurements we used laser pulses from an optical parametric oscillator (OPO) pump by a Ti:Sapphire mode-lock laser (Mira-HP, Coherent). The pulse duration is 250fs with a repetition rate of 78 MHz.

# Supplementary information

## S1. Device fabrication

Graphene/hBN heterostructure:

The hBN/graphene/hBN stacks were exfoliated and assembled using a van der Waals assembly technique and then transferred onto the top of the PCC cavity. The total thickness of the two BN layers is around 30 nm. Patterning the BN/Graphene/BN stack with hydrogen-silsesquioxane (HSQ) resist and $CHF_3+O_2$ plasma exposes the edges of graphene, which was subsequently contacted by Cr/Pd/Au (1/20/50 nm) metal leads using electron-beam evaporation. Typical device resistance as a function of gate voltage and device temperature is shown in Fig. S2 (a).

L3 photonic crystal nanocavity:

The L3 PCC is a widely used high Q photonic crystal nanocavity [18]. The PCC cavities were fabricated on a silicon-on-insulator wafer using a series of electron-beam lithography, reactive ion etching, and a wet-etch undercut of the insulator to produce free-standing membranes. The top silicon membrane has a thickness of ~220 nm with a PCC lattice period of a=440 nm and an air hole radius r=0.29a. A linear-three-hole (L3) defect in the middle of the PCC lattice serves to form confined optical cavity modes. Finite-difference time-domain simulations show the fundamental mode of the cavity resonant field (Fig. 1(c)).

The graphene layer interacts with the cavity through the evanescent field above the subwavelength membrane. The thickness of the membrane plus the thickness of the bottom layer of hBN define the magnitude of the evanescent electric field amplitude at the graphene location. From Lumerical simulations we have found it to be ~10% of the field maximum at the center of the structure.

## S2. Optical characterization

Scanning cross-polarized confocal microscope setup:

The fabricated PCC cavity was characterized in a scanning cross-polarized confocal microscope setup (Fig. S1 (a)) with broadband illumination (supercontinuum). The reflection from the cavity was collect and coupled to a spectrometer to analyze its spectrum (Fig. 2 (a)).

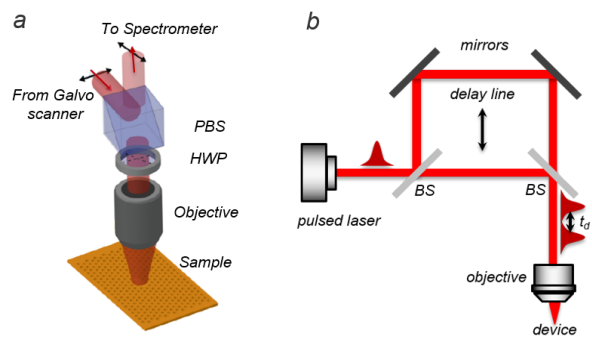

**Figure S1: Optical measurement setups.** (a) Schematics of the scanning cross-polarized confocal microscope setup. Here PBS is a polarizing beam splitter and HWP is a half-wave plate. (b) Schematics of the time-resolved optical measurement setup. The laser pulses are split by a beam splitter (BS) and recombined. One of the pulses travels on an adjustable length delay line and acquires a time delay $t_d$ with respect to the first puls.

Time-resolved optical measurements:
Laser pulses are generated from an optical parametric oscillator (OPO) pump by a Ti:Sapphire mode-lock laser (Mira-HP, Coherent). The pulse duration is 250fs with a repetition rate of 78 MHz. The input laser beam is split into two optical arms where one of the arms has a tunable delay relative to the other. The optical pulses were then combined and coupled onto the graphene bolometer via a microscope objective Fig. S1 (b). The JNP of the bolometer was measured while scanning the delay between the two input arms Fig. 4 (a).

Impedance matching of the RF circuit:
To achieve impedance matching of the RF circuit and the graphene device we use a LC resonant circuit. It defines a transmission band around a resonant frequency of ~70MHz (with typical values of other devices are in the range of ~50-200MHz) and a bandwidth of ~6MHz (typical values ~5-50MHz)[2,4] (Fig. S2 (b)). The challenge is to pick L and C components that allow for a good noise transmission of the JNP through the circuit for all resistances of the device $R(\Delta V_{bg}, T_s)$ (Fig. S2 (b) and (c)).

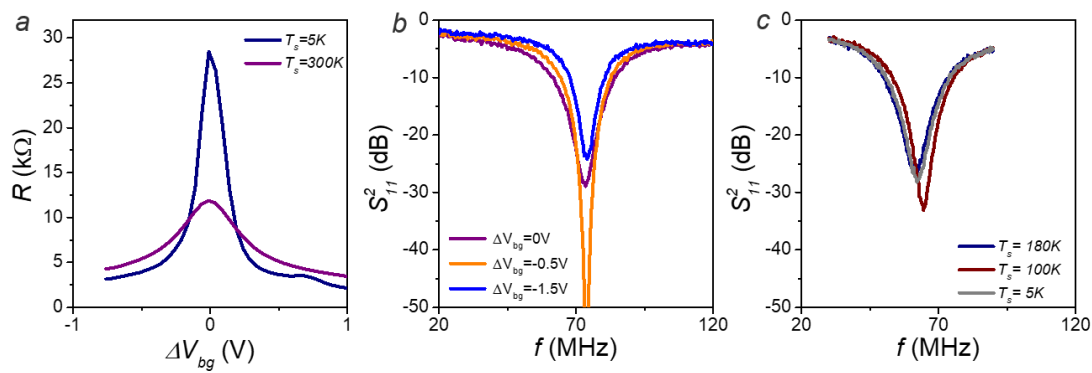

**Figure S2: Resistance and impedance matching of the device.** (a) Strong temperature dependence of the two-terminal device resistance R as a function of gate voltage $\Delta V_{bg}$ for different device temperatures $T_s$. (b) $S^2_{11}$ reflectance spectrum of the effective device LRC circuit for different gate voltages $\Delta V_{bg}$ at $T_s$=5K. (c) $S_{11}$ reflectance spectrum for various $T_s$ at $\Delta V_{bg}$=0V. (b) and (c) demonstrate good impedance matching for the whole parameter space.

Electronic temperature readout with JN:
Fig. S3 (a) shows a typical calibration curve for the JNP output voltage as a function of electronic temperature $T_e$. A linear fit to the JNP allows to calibrate $T_e$ of the electron gas in the graphene bolometer. As the device resistance R changes with gate voltage, it is necessary to calibrate $T_e$ for each gate voltage value separately. Generally, as from the intersection of the linear JNP calibration curve with $T_e$=0K, the JNP signal is not zero, this residual signal defines the temperature independent system noise $T_{sys}$, which is usually dominated by the amplifier noise. It is common to quantify $T_{sys}$ by the intersection of the linear fit in Fig. S3 (a) with JNP=0V, where we extract a $T_{sys}$ ~ 78K. The so extracted $T_{sys}$ is slightly bigger than when calculated from the Dicke formula, but is in very good agreement with previous experiments, which used equivalent circuits[4]. We suspect that this slight discrepancy might be attributed to an effectively larger transmission bandwidth, than extracted from the FWHM of the $S^2_{11}$ reflectance spectra in Fig. S2 (b).

Measurement sensitivity:
In order to estimate the measurement sensitivity we perform more than 1000 independent $T_e$ measurements at a fixed $T_s$, where we find a measurement uncertainty of $\delta T_e \sim 10$mK for $T_s$=5K (Fig. S3 (b)) (and $\delta T_e \sim 18$mK for $T_s$=100K, $\delta T_e \sim 40$mK for $T_s$=300K). Also at $T_s$=5K we can fit $\delta T_e$ as a function of the measurement integration time $t_{in}$ with the Dicke radiometer relation (Fig. S3 (c)), and find excellent agreement with the expected behaviour.

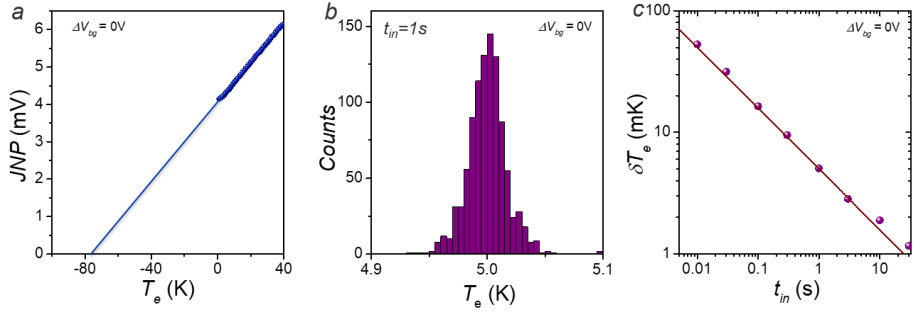

**Figure S3: JN thermometry calibration and sensitivity.** (a) The integrated device noise output JNP is proportional to $T_e$. The slope of the JNP($T_e$) dependence allows to calibrate the thermometer. (b) Histogram of independent $T_e$ measurements using an integration time of $t_{in}$=1s, which gives a measurement uncertainty of $\delta T_e \sim$ 10mK, at $T_s$=5K. (c) The measurement uncertainty $\delta T_e$ (bullet points) follows the Dicke relation (fit) for different integration times $t_{in}$. From comparison with the experiment, where throughout we used a $t_{in}$ =1s, we find that the sensitivity in our measurement is significantly limited by $T_{sys}$ at low temperatures.

### S4. Numerical modeling of Johnson noise temperature
In order to describe the Johnson noise analytically, we begin from the assumption of a well-defined local electron temperature $T_e(r)$ at a given position r in the sample. This assumption is equivalent to assuming that the electron mean free path (resulting, for example, from scattering by disorder impurities) is much shorter than any of the geometric dimensions of the sample. In this description, each differential area of the sample effectively acts as its own independent noise source and contributes independently to the total Johnson noise measured at the contacts. In particular, the noise current IN can be written

$$I_N = \int [j_N(r) \cdot \nabla \phi(r)] d^2 r , \text{(S.1)}$$

where $j_N(r)$ is the local noise current and $\phi(r)$ is a weighting function that obeys the same Laplace equation as the source-drain current. In our problem, $\phi(r)$ is equivalent to the electric potential that results when a unit voltage is applied to the source contact and the drain contact is grounded. Fluctuations to the noise current add in quadrature with each other, $I_N^2 = \int \langle j_N^2(r)\rangle d^2r$, and since for Johnson noise $\langle j_N^2(r)\rangle \propto T_e(r)$ we arrive at a simple expression for the Johnson noise temperature:

$$T_{JN} = \frac{\int T_e(r)(\nabla\phi)^2 d^2r}{\int (\nabla\phi)^2 d^2r} \quad .\text{(S.2)}$$

For our device, the function $\phi(r)$ and its spatial gradient can be found numerically using a numeric partial differential equation solver (MATLAB) applied to the device geometry (Fig. S4 (a)).

The spatially-varying electron temperature is described by the two-dimensional heat equation

$$\dot{q}(r) = -\nabla \cdot [\kappa(r)\nabla T_e(r)] + \Sigma_{ep}[T_e^\delta - T_s^\delta] \, ,\text{(S.3)}$$

where $\dot{q}(r)$ is the power per unit area absorbed by the electrons at position $r$, $\kappa(r)$ is the spatially-varying thermal conductivity, $\Sigma_{ep}$ is the electron-phonon coupling constant, $T_s$ is the phonon (device) temperature, and $\delta$ is an exponent that describes the electron-phonon coupling and is typically equal to 3. The first term on the right-hand side describes diffusive motion of the electrons, which allows heat to dissipate by diffusion into the contacts, and the second term describes heat loss to phonons. The electron thermal conductivity is given by the Wiedemann-Franz law,

$$\kappa(r) = \frac{\pi^2}{3}\left(\frac{k}{e}\right)\sigma T_e(r) \, .\text{(S.4)}$$

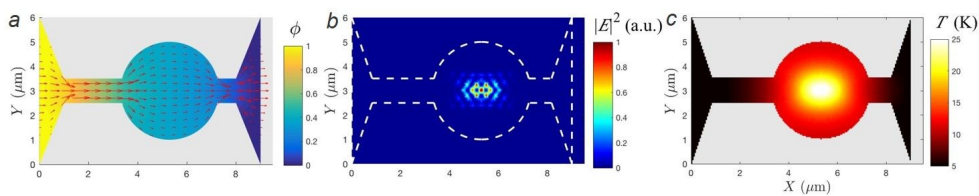

**Figure S4: Numeric evaluation of the Johnson noise temperature.** (a) The weighting function $\phi(r)$, which is given by a solution to the Laplace equation with the source set to $\phi = 1$ and the drain set to $\phi = 0$. Red arrows show the gradient $-\vec{\nabla}\phi(r)$. (b) The intensity of the cavity mode as a function of position. The relative position of the graphene sample is illustrated with dashed white lines. (c) Numerical solution for the spatially-varying electron temperature $T_e(r)$, assuming an absorbed power $P$=20 nW and a base temperature $T_s$= 5 K.

The heating of electrons comes predominantly from the cavity mode, so that $\dot{q}(r)$ is proportional to the squared electric field of the cavity. We therefore write $\dot{q}(r) = P f_c(r)$, where $P$ is the total power absorbed and $f_c(r)$ is a distribution function that is proportional to the squared electric field in the cavity mode and which integrates to unity, $\int f_c(r)d^2r = 1$. The cavity mode profile $f_c(r)$ is calculated numerically using Lumerical, and is plotted in Fig. S3 (b). In principle, there is a small amount of additional power that is absorbed directly by graphene electrons from the laser spot, independently of the cavity mode. This additional absorption is small compared to the absorption from the cavity mode, and does not noticeably alter our results. In principle, however, it can be added to the function $\dot{q}(r)$.

If the electron-phonon coupling constant $\Sigma_{ep}$ is known, then we can solve Eqn. (S.3) numerically to determine the electron temperature $T_e(r)$ as a function of position. This is done numerically, with the boundary conditions $T_e=T_s$ the contacts (which are good heat absorbers due to their high electronic density of states) and reflecting boundary conditions at the other edges of the sample. An example for the solution $T_e(r)$ is shown in Fig. S4 (c). Once $T_e(r)$ is known, the Johnson Noise temperature $T_{JN}$ can be calculated using equation (S.2).

The value of the constant $\Sigma_{ep}$ for our device is not known *a priori*. We infer $\Sigma_{ep}$ by fitting our measured data for $T_{JN}$ versus $P$, as shown in Fig. 3 (a) of the main text, which gives $\Sigma_{ep} \approx 0.038$ and $\delta \approx 3.0$.

Finally, we can also use our numerical modeling to describe the variation in the Johnson noise temperature as a function of the position of the laser spot. For this calculation we assume that the power in the cavity mode is proportional to the laser power multiplied by the spatial overlap between the laser spot profile and the cavity mode profile. That is,

$$\dot{q}(r) = P_0 f_c(r) \int f_c(r') f_l(r') d^2 r', \quad (S.5)$$

where $P_0$ is proportional to the laser power and $f_l(r)$ is a Gaussian distribution that describes the laser spot intensity. In our modeling we assume a beam spot width 1.5 $\mu m$. An example calculation of the resulting Johnson noise $T_{JN}$ as a function of the beam spot position is given in Fig. S4.

Equation (S.5) can be justified theoretically by considering a generalization of the Breit-Wigner model for coupling of a continuous excitation (here, light) to a resonant excitation (here, the photonic cavity mode). In particular, if light with energy $\omega$ is coupled to a resonant excitation mode with energy $\omega_0$ and damping $\gamma$, then the absorption coefficient

$$A = \frac{4\Gamma\gamma}{(\omega-\omega_0)^2+(\Gamma+\gamma)^2}. \quad (S.6)$$

Here, $\Gamma$ represents the resonance width, and in our problem the damping $\gamma$ arises from coupling between the cavity mode and the graphene electrons. In the traditional Breit-Wigner problem, $\Gamma = \nu g^2$, where $\nu$ is the density of states for the continuous excitation and *g* is the coupling constant between the resonant mode and the free photons. In our problem, the coupling constant $g(r)$ at a given spatial location *r* is proportional to the magnitude of the local electric field $E(r)$ in the cavity mode. By Eq. (S.6), we therefore have a local absorption coefficient *A(r)* that is proportional to the square of the cavity mode electric field, $A(r) \propto |E(r)|^2 \propto f_c(r)$.

We arrive at Eq. (S.5) by defining the total power in the cavity mode as the integral of the local laser power $P(r) \propto f_l(r)$ multiplied by the local absorption coefficient *A(r)*. One can note that in doing so we ignore the potential coupling between free photon modes emitted at different spatial locations of the sample. Such coupling can be expected to be small so long as the integrated absorption coefficient is small compared to unity.